\documentclass[%
reprint,longbibliography,
 amsmath,amssymb,
 aps,
]{revtex4-2}

\usepackage{bm}
\usepackage{graphicx}
\usepackage{enumitem}   
\usepackage{cleveref}
\graphicspath{{graphics/}}
\usepackage{float}

\begin{document}

\widetext

\title{First hitting times between a run-and-tumble particle and a stochastically-gated target}

\author{Gabriel Mercado-V\'asquez}%
\email{gabrielmv.fisica@gmail.com}
\author{Denis Boyer}%
 \email{boyer@fisica.unam.mx}
\affiliation{%
 Instituto de F\'isica, Universidad Nacional Aut\'onoma de M\'exico, Mexico City 04510, Mexico
}%
\date{\today}

\begin{abstract}
We study the first hitting time statistics between a one-dimensional run-and-tumble particle and a target site that switches intermittently between visible and invisible phases. The two-state dynamics of the target is independent of the motion of the particle, which can be absorbed by the target only in its visible phase. We obtain the mean first hitting time when the motion takes place in a finite domain with reflecting boundaries. Considering the turning rate of the particle as a tuning parameter, we find that ballistic motion  represents the best strategy to minimize the mean first hitting time. However, the relative fluctuations of the first hitting time are large and exhibit non-monotonous behaviours with respect to the turning rate or the target transition rates. Paradoxically, these fluctuations can be the largest for targets that are visible most of the time, and not for those that are mostly invisible or rapidly transiting between the two states. On the infinite line, the classical asymptotic behaviour $\propto t^{-3/2}$ of the first hitting time distribution is typically preceded, due to target intermittency, by an intermediate scaling regime varying as $t^{-1/2}$. The extent of this transient regime becomes very long when the target is most of the time invisible, especially at low turning rates. In both finite and infinite geometries, we draw analogies with partial absorption problems.
\end{abstract}

\maketitle

\section{Introduction}
Looking for an object without having any information about its location is, in most cases, an arduous task that is often limited by the strategy adopted by the searcher\cite{Luz_2009}. Furthermore, if the state of the object fluctuates in time in such a way that it is not always visible or detectable, the search task might become extremely difficult. Viewing the search problem as a diffusion-reaction process, one may seek to determine the time needed for a particle to react for the first time with one or several targets that change stochastically between reactive and unreactive states. 

Search processes in fluctuating environments are widely present in nature. In the context of foraging ecology, some organisms adopt crypsis as their primary defense and develop a wide range of mechanisms to become less detectable for some periods of time and thus prevent predation\cite{ruxton2004avoiding,stevens2009animal,edmunds1990evolution,gendron1983searching,o1990search}. At the molecular level, the binding of ligands to proteins depends on the conformational state of one of the two components; in some cases, different side chains of the protein serve as gates that block the entrance of ligands, and fluctuations in protein reactivity can be successfully described in terms of stochastic gating processes\cite{mccammon1981gated,szabo1982stochastically,zhou1996theory}. Reactions in imperfect media are often modelled by an absorption probability less than unity, such that a diffusive particle may react after several passages over the target region\cite{ben1993partial,redner2001guide,grebenkov2019imperfect}. A similar situation arises for the first detected passage of quantum walkers, since in these systems the trajectory of a particle cannot be continuously recorded and a first arrival time at a given threshold cannot be defined. Instead, measurements at the threshold site are sampled at a finite rate until the particle is detected\cite{Barkai2017,BarkaiPhysRev2018}. 

The study of the first passage properties of systems with perfectly absorbing  targets has been gaining more attention in the past few decades. Special interest has been dedicated to non-Brownian search processes such as L\'evy flights\cite{viswanathan1999optimizing,viswanathan2002levy,viswanathan2011physics}, ballistic movements \cite{viswanathan1999optimizing,james2008optimizing}, run-and-tumble motion \cite{angelani2014first,Malakar_2018}, or types of motion that combine diffusive phases with ballistic relocations\cite{benichou2005optimal,benichou2011intermittent,benichou2006two}.  This growing interest lies in the fact that many organisms, ranging from motile cells to animals, perform non-Brownian motions in order to efficiently explore their environment \cite{harris2012generalized,deJager1551,sims2008scaling}.

The non-Brownian processes of interest here is the run-and-tumble (RT) motion, which consists of straight-line motion at constant speed (run) interspersed with random re-orientations occurring with  a constant rate (tumble). Despite of its simplicity, the RT motion has served to model a wide range of non-equilibrium systems such as self-propelled particles\cite{elgeti2015run,Malakar_2018}, electron collisions in a Lorentz gas\cite{martens2012probability} or the motion of bacteria such as \textit{E. coli}, \textit{Salmonella}  or the marine bacteria \textit{P. haloplanktis } \cite{berg2008coli,stocker2011reverse}. The motility of microswimers in diverse environments has been better understood thanks to analytical results on the RT model. In confined environments, run-and-tumble particles (RTP) tend to accumulate near the boundaries\cite{Malakar_2018,elgeti2015run}. In the presence of steady potentials, transitions between active and passive-like behaviour have been observed \cite{Abhishek2019}. RTPs subject to periodic and asymmetric potentials exhibit a so-called active ratchet effect\cite{Angelani_2011}.  In \cite{angelani2015run} a run-and-tumble particle in the presence of imperfect boundaries has been considered. These studies have brought evidence that run-and-tumble particles exhibit properties that contrast with those of their Brownian counterparts.

First passage problems with non-Brownian particles in the presence of perfectly absorbing or reflecting boundaries have been well studied over the recent years\cite{weiss2002some,PhysRevE.71.012101,metzler2000boundary,rangarajan2000first,angelani2014first,Malakar_2018}, but less attention has been paid to time-dependent environments. A realistic description of biological systems must contemplate the presence of external noises that can interrupt or affect the interaction between the components of a system. For instance, in gene expression, proteins alternate between 3$d$ diffusion within the cell and $1d$ sliding along the DNA strands when they search for their specific binding sites\cite{kong2017rad4,benichou2011intermittent}. Different kinds of  motions  have been proposed for the sliding phase of proteins, from simple diffusion to ballistic\cite{eliazar2007searching}. Nevertheless, the conformational chromatin structure or regulatory molecules may temporarily prevent protein binding on their target sites\cite{elowitz2002stochastic,munsky2012}. In another context, marine bacteria must navigate poor environments where nutrients are typically found in small patches that can be seen as point sources of food\cite{Blackburn2254}. Due to physical processes that dissipates the nutrients, these patches have a short life, lasting for tens of seconds, which forces bacteria to respond quickly to the formation of patches\cite{Stocker4209}.

In this work, we study the first hitting properties of a one-dimensional run-and-tumble particle searching for a gated target, {\it i.e.}, a site that stochastically switches between two conformations: an \lq \lq active\rq\rq\ state that absorbs the RTP upon encounter, and an inactive state in which the target is \lq\lq invisible\rq\rq\ and cannot be detected during a passage. By considering this system, we also extend recent results obtained for a Brownian particle\cite{PRLFirstHittingTimes}. Here we analyze the RT motion in several geometries: $(i)$ the finite domain, with the particle bounded by two reflective walls symmetrically placed around the target, $(ii)$ the infinite line, as a special case in which the walls are infinitely far away. In the first case, quantities of primary interest are the mean first hitting time (MFHT) and the standard deviation of the hitting time around this mean. We find  that the global MFHT averaged over the particle initial positions is always minimal for a ballistic particle, whereas less persistent searchers are suboptimal. However, the relative variance of the first hitting time can take very large values and exhibits a non-monotonic behaviour with respect to the parameters that control the intermittent dynamics of the target. In the unbounded case, the MFHT diverges and we study the asymptotic behaviour of the full first hitting time distribution (FHTD). The decay of the distribution at large hitting times $t$ is decomposed into two scaling regimes: a $t^{-3/2}$ asymptotic regime that is characteristic of unbiased motion, and an intermediate regime varying as $t^{-1/2}$, which represents a much slower decay and whose range can be varied depending on the target and particle rates. This intermediate regime originates from the gating dynamics of the target and is quite generic: it was first elucidated in the case of Brownian motion \cite{PRLFirstHittingTimes} and further observed recently in the context of random searches on networks \cite{scher2021unified}.

The paper is organized as follows: in Section  \ref{finiteDomain} we begin by introducing the model and deduce the governing equations for the survival probabilities, which are solved in the Laplace domain in the general case. With the help of these results, in Section \ref{MFHT} we compute the MFHT and analyze this quantity for several limiting cases of the intermittent dynamic parameters and of the particle motion. Subsequently, in Section \ref{SecRelVar} we calculate the relative variance of the first hitting times. Section \ref{InfiniteDomain} is devoted to analyzing the FHTD in the unbounded case. Finally, we discuss our findings in Section \ref{Discussion}.

\section{General setup and solution}\label{finiteDomain}

We start by defining a time-dependent binary variable $\sigma(t)$ that describes the state of a target placed at the origin of a one-dimensional space, and that takes the value $\sigma=1$ when it is visible (or active) and $\sigma=0$ when it is hidden (or inactive). The target state switches at exponentially distributed times, with rate $a$ for the transition $0\to 1$ and with rate $b$ for the transition $1\to 0$. Therefore, the mean duration of the active (inactive) phase is $1/b$ ($1/a$, respectively) and the overall probability to find the target in the active state is $a/(a+b)$. We then consider a run-and-tumble particle that moves at constant speed $v$ and changes its direction at a rate $\gamma$. Thus, the motion is governed by the equation
\begin{equation}
    \frac{dx}{dt}=\Gamma(t)\label{LangevinRT}
\end{equation}
where $\Gamma(t)$ is a dichotomous noise which takes constant values, $+v$ or $-v$, during exponentially distributed time intervals of mean duration $1/\gamma$. Two reflective barriers are placed at the positions $\pm L$, constraining the movement of the particle to the interval $(-L,L)$. 

\begin{figure}[htp]
\centering
			\includegraphics[width=.48\textwidth]{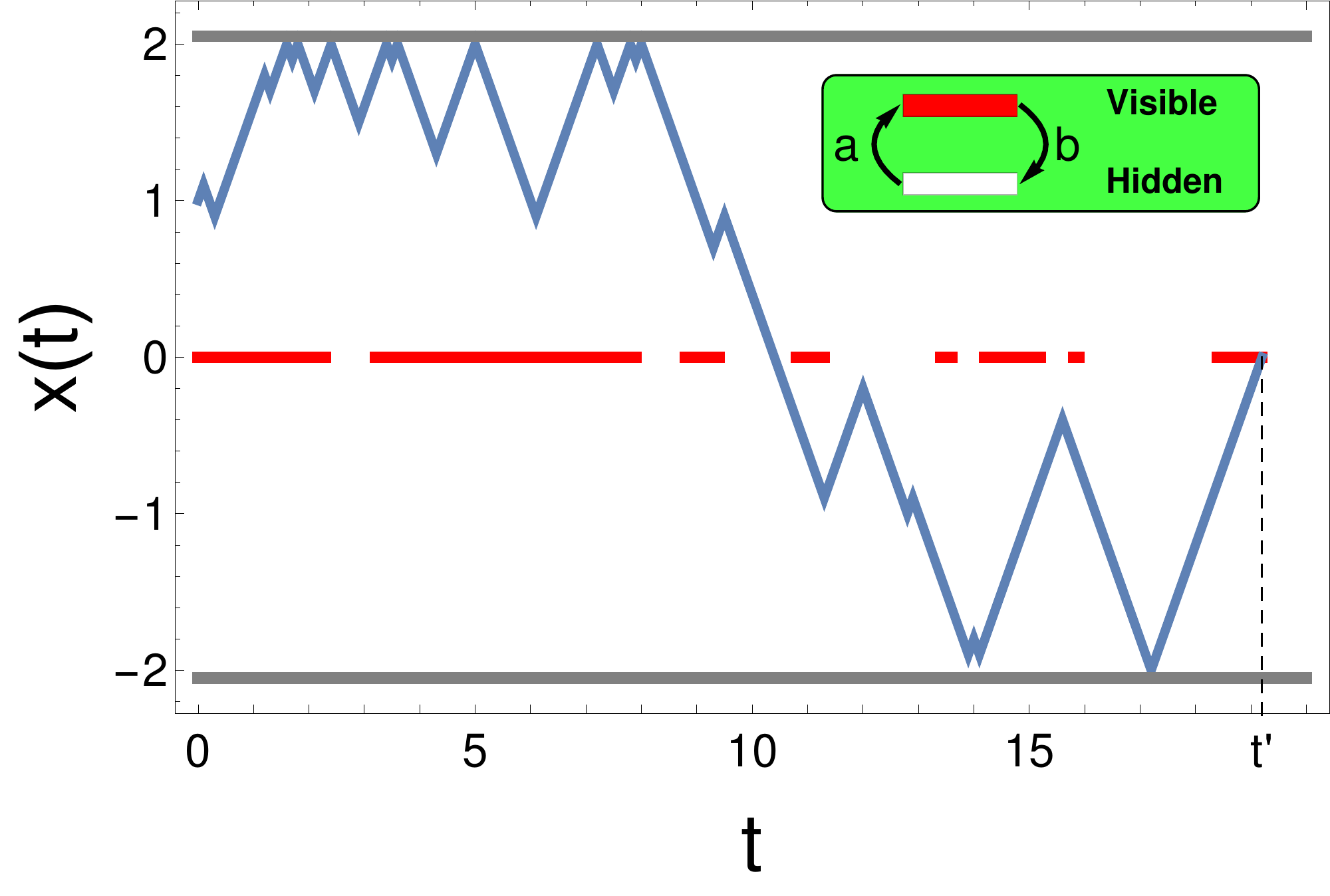}
            
			\caption{Run-and-tumble motion (blue line) in 1D in the presence of an intermittent target placed at the origin. The periods of the target in the visible state are represented by red segments, whereas the white intervals represent the target in its hidden state.  The gray lines represent the reflective boundaries located at $L$ and $-L$, with $L=2$.}
			\label{fig:walk1DRT}
\end{figure}

Fig. \ref{fig:walk1DRT} depicts an example of first encounter in which the RTP (blue line) starts from $x(t=0)=L/2$ and with the initial velocity $+v$, whereas the target is initially in the visible state (in red). During the search process, the particle reverses its direction of motion randomly through tumble events  or by reflections against the walls. If the particle crosses the origin while the target is hidden (white line), there is no encounter and the RTP simply follows its way; however, if the target is visible, it is detected and the process ends at the first hitting/encounter time $t'$. 

In order to analyze the statistics of $t'$ in this system, let us denote as $Q^+_{0}(x,t)$ [and $Q^-_{0}(x,t)$] the probability that the particle has not reacted up to time $t$, given the initial target state $\sigma(t=0)=0$, the initial particle position $x\in(-L,L)$ and the initial velocity $+v$ [$-v$, respectively]. Similarly, we denote as $Q^+_{1}(x,t)$ and $Q^-_{1}(x,t)$ the survival probabilities when the target initial state is $\sigma(t=0)=1$. We show in Appendix \ref{App_BFP} that these four probabilities satisfy the backward Fokker-Planck equations
\begin{equation}
\begin{aligned}
    \frac{\partial Q_0^+}{\partial t}&=&v\frac{\partial Q_0^+}{\partial x}-\gamma(Q_0^{+}-Q_0^{-})-a(Q_0^{+}-Q_1^{+}),\\
     \frac{\partial Q_0^-}{\partial t}&=&-v\frac{\partial Q_0^{-}}{\partial x}-\gamma(Q_0^--Q_0^+)-a(Q_0^{-}-Q_1^{-}),\\
      \frac{\partial Q_1^+}{\partial t}&=&v\frac{\partial Q_1^+}{\partial x}-\gamma(Q_1^+-Q_1^-)-b(Q_1^+-Q_0^+),\\
      \frac{\partial Q_1^-}{\partial t}&=&-v\frac{\partial Q_1^-}{\partial x}-\gamma(Q_1^--Q_1^+)-b(Q_1^--Q_0^-).
      \label{SPsystem1}
      \end{aligned}
\end{equation}
For the initial particle position $x>0$, the system \eqref{SPsystem1} will satisfy the following boundary conditions:
\begin{align}
      &Q^-_1(x=0^+,t)=0,\label{BC1runtumble}\\
      &Q^+_0(x=0,t)=Q^-_0(x=0,t),\label{BC2runtumble}\\
     &Q^+_0(x=L,t)=Q^-_0(x=L,t),\label{BC3runtumble}\\
    &Q^+_1(x=L,t)=Q^-_1(x=L,t).\label{BC4runtumble}
\end{align}
The first condition asserts that the particle detects the target in state $\sigma=1$ when it goes leftward from its immediate vicinity on the right. Eq. \eqref{BC2runtumble} stems from symmetry, as the target is placed in the middle of the domain. Eqs. \eqref{BC3runtumble}-\eqref{BC4runtumble} set the reflective condition on the wall placed at $x=L$.

We can also average over the initial target states and assume equal probabilities for the initial positive and negative particle velocities. The resulting average survival probability is thus:
\begin{align}
Q_{av}(x,t)=&\frac{b}{a+b}\left(\frac{Q^+_0(x,t)+Q^-_0(x,t)}{2}\right)\nonumber\\
&+\frac{a}{a+b}\left(\frac{Q^+_1(x,t)+Q^-_1(x,t)}{2}\right).\label{qavRT}
\end{align}

The probability distributions of the first hitting time ($t$) are denoted as $P^{+}_0(x,t)$, $P^{-}_0(x,t)$, $P^{+}_1(x,t)$ and $P^{-}_0(x,t)$ with the same notations as before for the initial conditions. These first hitting time densities (FHTD) are deduced from the survival probabilities through the usual relation\cite{redner2001guide}:
\begin{equation}
P^{\pm}_{0,1}(x,t)=-\frac{\partial Q^{\pm}_{0,1}(x,t)}{\partial t}.\label{fhtd}
\end{equation}

In the following, it is convenient to set $\gamma^{-1}$ as the unit of time and $v\gamma^{-1}$ as the unit of length, so that the system \eqref{SPsystem1} can be recast, in the dimensionless variables $\tau=t\gamma$ and $z=x\gamma/v$, as
\begin{equation}
\begin{aligned}
    \frac{\partial Q_0^+}{\partial \tau}&=&\frac{\partial Q_0^+}{\partial z}-(Q_0^{+}-Q_0^{-})-\alpha(Q_0^{+}-Q_1^{+}),\\
     \frac{\partial Q_0^-}{\partial \tau}&=&-\frac{\partial Q_0^{-}}{\partial z}-(Q_0^--Q_0^+)-\alpha(Q_0^{-}-Q_1^{-}),\\
      \frac{\partial Q_1^+}{\partial \tau}&=&\frac{\partial Q_1^+}{\partial z}-(Q_1^+-Q_1^-)-\beta(Q_1^+-Q_0^+),\\
      \frac{\partial Q_1^-}{\partial \tau}&=&-\frac{\partial Q_1^-}{\partial z}-(Q_1^--Q_1^+)-\beta(Q_1^--Q_0^-).
      \label{SPsystem2}
      \end{aligned}
\end{equation}
which only depends on two parameters, the dimensionless rates: 
\begin{align}
\alpha=a\gamma^{-1},\\
\beta=b\gamma^{-1}.
\end{align}
In these dimensionless units, the particle is thus restricted to $z\in(-\ell,\ell)$, where $\ell$ is the re-scaled domain size 
\begin{equation}
    \ell=L\gamma/v.
\end{equation}

Introducing the Laplace transform $\widetilde{Q}(z,s)=\int^\infty_0e^{-s\tau}Q(z,\tau) d\tau$ and using the initial condition $Q(z,\tau=0)=1$ for $0<z<\ell$, the Laplace transform of  Eqs. \eqref{SPsystem2} gives
\begin{equation}
    \frac{\partial}{\partial z}\begin{pmatrix} \widetilde{Q}_0^+\\ \widetilde{Q}_0^-\\ \widetilde{Q}_1^+\\ \widetilde{Q}_1^- 
    \end{pmatrix}=\mathbb{A}\begin{pmatrix} \widetilde{Q}_0^+\\ \widetilde{Q}_0^-\\ \widetilde{Q}_1^+\\ \widetilde{Q}_1^- 
    \end{pmatrix}-\begin{pmatrix} 1\\ -1\\ 1\\ -1    \end{pmatrix}\label{SPsystem3}
\end{equation}
where
\begin{equation}
    \mathbb{A}=\begin{pmatrix} 1+s+\alpha & -1 & -\alpha & 0 \\
    1 & -1-s-\alpha & 0 & \alpha \\
    -\beta & 0 & 1+s+\beta & -1 \\
    0 & \beta & 1 & -1-s-\beta
    \end{pmatrix}\label{matrixA}
\end{equation}

The homogeneous part of Eq. \eqref{SPsystem3} can be solved with the ansatz $\boldsymbol{\xi} e^{ \lambda z}$, where $\lambda$ and the vector $\boldsymbol{\xi}$ must be determined by diagonalizing \eqref{matrixA}, whereas the inhomogeneous solution is simply $\widetilde{Q}^{\pm}_{0,1}=1/s$. After straightforward algebra,  the general solution  $\mathbf{\widetilde{Q}}=\begin{pmatrix}
\widetilde{Q}^+_0 & \widetilde{Q}^-_0 & \widetilde{Q}^+_1 & \widetilde{Q}^-_1\end{pmatrix}^T$ is given by the following linear combination of terms
\begin{equation}
\mathbf{\widetilde{Q}}=A_1\boldsymbol{\xi}_1e^{-\lambda_1 z}+A_2\boldsymbol{\xi}_2e^{\lambda_1 z}+A_3\boldsymbol{\xi}_3e^{-\lambda_2 z}+A_4\boldsymbol{\xi}_4e^{\lambda_2 z}+\mathbf{\widetilde{Q}}_{inh},\label{gralsolruntumble}    
\end{equation}
where the factors $A_k$ are determined by the boundary conditions and  $\mathbf{\widetilde{Q}}_{inh}$ is the inhomogeneous solution with each entry equal to $1/s$. In Eq. \eqref{gralsolruntumble}, the eigenvalues $\lambda_1$ and $\lambda_2$ are positive and given by
\begin{equation}
    \lambda_{1}=\sqrt{s}\sqrt{2+s},\ \    \lambda_{2}=\sqrt{\alpha+\beta+s}\sqrt{2+\alpha+\beta+s},
\end{equation}
whereas the eingenvectors are
\begin{align*}
&\boldsymbol{\xi}_1=\begin{pmatrix} 1+s-\lambda_1 \\ 1 \\ 1+s-\lambda_1\\ 1 \end{pmatrix},  \boldsymbol{\xi}_2=\begin{pmatrix} 1+s+\lambda_1 \\ 1 \\ 1+s+\lambda_1\\ 1 \end{pmatrix}, \\
&\boldsymbol{\xi}_3=\begin{pmatrix} -\frac{\alpha\left(1+s+\alpha+\beta-\lambda_2\right)}{\beta} \\ -\frac{\alpha}{\beta} \\ 1+s+\alpha+\beta-\lambda_2\\ 1 \end{pmatrix}, \boldsymbol{\xi}_4=\begin{pmatrix} -\frac{\alpha\left(1+s+\alpha+\beta+\lambda_2\right)}{\beta} \\ -\frac{\alpha}{\beta} \\ 1+s+\alpha+\beta+\lambda_2\\ 1 \end{pmatrix}.
\end{align*}
From Eq. (\ref{qavRT}), the average survival probability takes a simpler form:
\begin{equation}
    \widetilde{Q}_{av}(z,s)=\frac{s+2-\lambda_1}{2}A_1e^{-\lambda_1 z}+\frac{s+2+\lambda_1}{2}A_2e^{\lambda_1 z}+\frac{1}{s}\label{meanQ}.
\end{equation}
The boundary conditions in the Laplace domain are recast as
\begin{align}
      \widetilde{Q}^-_1(z=0^+,s)=0,\label{BC1runtumbleS}\\
      \widetilde{Q}^+_0(z=0,s)=\widetilde{Q}^-_0(z=0,s),\label{BC2runtumbleS}\\
     \widetilde{Q}^+_0(z=\ell,s)=\widetilde{Q}^-_0(z=\ell,s),\label{BC3runtumbleS}\\
    \widetilde{Q}^+_1(z=\ell,s)=\widetilde{Q}^-_1(z=\ell,s).\label{BC4runtumbleS}
\end{align}
 With these conditions, the general solution in Eq. \eqref{gralsolruntumble} admits a unique solution with the factors $A_k$ given by
\begin{align}
   A_1&=-\frac{e^{\lambda _1 \ell} \left(s+\lambda _1\right) \text{csch}\left(\lambda _1 \ell\right)}{2 s \left(\lambda _1 \coth \left(\lambda _1 l\right)+\frac{s}{\alpha }
   \left(\alpha +\beta +\frac{\beta  \lambda _2 \coth \left(\lambda _2 \ell\right)}{\alpha +\beta +s}\right)\right)},\\
   A_2&=\frac{e^{-\lambda _1 \ell} \left(s-\lambda _1\right) \text{csch}\left(\lambda _1 \ell\right)}{2 s \left(\lambda _1 \coth \left(\lambda _1 l\right)+\frac{s}{\alpha }
   \left(\alpha +\beta +\frac{\beta  \lambda _2 \coth \left(\lambda _2 \ell\right)}{\alpha +\beta +s}\right)\right)},\\
   A_3&=-\frac{\beta e^{\lambda _2 \ell} \left(s+\alpha+\beta+\lambda _2\right)(s+\alpha+\beta)^{-1} \text{csch}\left(\lambda _2 \ell\right)}{2\alpha \sqrt{s} \left(\lambda _1 \coth \left(\lambda _1 l\right)+\frac{s}{\alpha }
   \left(\alpha +\beta +\frac{\beta  \lambda _2 \coth \left(\lambda _2 \ell\right)}{\alpha +\beta +s}\right)\right)},\\
   A_4&=\frac{\beta e^{-\lambda _2 \ell} \left(s+\alpha+\beta-\lambda _2\right)(s+\alpha+\beta)^{-1} \text{csch}\left(\lambda _2 \ell\right)}{2\alpha \sqrt{s} \left(\lambda _1 \coth \left(\lambda _1 l\right)+\frac{s}{\alpha }
   \left(\alpha +\beta +\frac{\beta  \lambda _2 \coth \left(\lambda _2 \ell\right)}{\alpha +\beta +s}\right)\right)}.
\end{align}
Inserting these expressions into the average survival probability \eqref{meanQ} yields
\begin{equation}
\begin{aligned}
\widetilde{Q}_{av}(z,s)&=\frac{1}{s}-\frac{\lambda_1\text{csch}\left(\lambda _1 \ell\right)\cosh\left(\lambda_1(z-\ell)\right)}{s\lambda_1\coth{\lambda_1\ell}+\frac{s^2}{\alpha}\left(\alpha+\beta+\frac{\beta  \lambda _2 \coth \left(\lambda _2 \ell\right)}{\alpha +\beta +s}\right)}.
\end{aligned}\label{meanQ2}
\end{equation}
The FHTD can be obtained from relation \eqref{fhtd} and the fact that, in the Laplace domain, $\partial Q(z,\tau)/\partial \tau$ transforms into $-1+s\widetilde{Q}(z,s)$. Therefore, we have
\begin{equation}
\begin{aligned}
\widetilde{P}_{av}(z,s)&=\frac{\lambda_1\text{csch}\left(\lambda _1 \ell\right)\cosh\left(\lambda_1(z-\ell)\right)}{\lambda_1\coth{\lambda_1\ell}+\frac{s}{\alpha}\left(\alpha+\beta+\frac{\beta  \lambda _2 \coth \left(\lambda _2 \ell\right)}{\alpha +\beta +s}\right)}
\end{aligned}\label{meanP2}.
\end{equation}

Seeking for an inversion of Eqs. \eqref{meanQ2} or \eqref{meanP2} does not look as simple as we would wish. However, one can exactly obtain from the above solution the mean first hitting time (MFHT), the second moment of the first hitting time distribution, as well as the behaviors of the tails of the full distribution.  In the following sections we calculate the MFHT and the variance. We leave for section \ref{InfiniteDomain} the analysis of the FHTD in the limit  $\ell\rightarrow\infty$.



\section{Mean first hitting time}\label{MFHT}

The mean first hitting time is given by $t_1(x)=\int_0^\infty t P_{av}(x,t)dt$, which, in units of $\gamma^{-1}$, can be written as $\tau_1(z)=\int_0^\infty \tau P_{av}(z,\tau)d\tau$. It is obtained from the Laplace transform of the survival probability if we use the relation (\ref{fhtd}) and integrate by parts to get $\tau_1(z)= \widetilde{Q}_{av}(z,s=0)$. Evaluating Eq. (\ref{meanQ2}) in the limit $s\rightarrow0$, we thus obtain the dimensionless MFHT 
\begin{align}
   \tau_1(z)=&(2\ell-z)z+\ell+\frac{\beta\ell}{ \alpha}\nonumber\\
   &+\frac{\beta\ell}{ \alpha}\sqrt{\frac{\alpha+\beta+2}
    {\alpha+\beta}}\coth{\ell\sqrt{\alpha+\beta}\sqrt{\alpha+\beta+2}}\label{MeanTimeDimless}.
\end{align}

It is convenient to define the global MFHT (in units of $\gamma^{-1}$), $\tau_G$, which is obtained by averaging $\tau_1(z)$ over $z$, which corresponds to having a uniform distribution of starting positions:
\begin{equation}
    \tau_G=\frac{1}{\ell}\int_0^{\ell}\tau_1(z)dz,\label{DeftauG}
\end{equation}
or, from Eq. \eqref{DeftauG},
\begin{align}
    \tau_G=&\ell+\frac{2}{3}\ell^2+\frac{\beta\ell}{\alpha}\nonumber\\
    &+\frac{\beta\ell}{\alpha}\sqrt{\frac{\alpha+\beta+2}
    {\alpha+\beta}}\coth{\ell\sqrt{\alpha+\beta}\sqrt{\alpha+\beta+2}}.\label{GMeanTimeDimless}
\end{align}

\begin{figure}[htbp]
\centering
      \includegraphics[width=.48\textwidth]{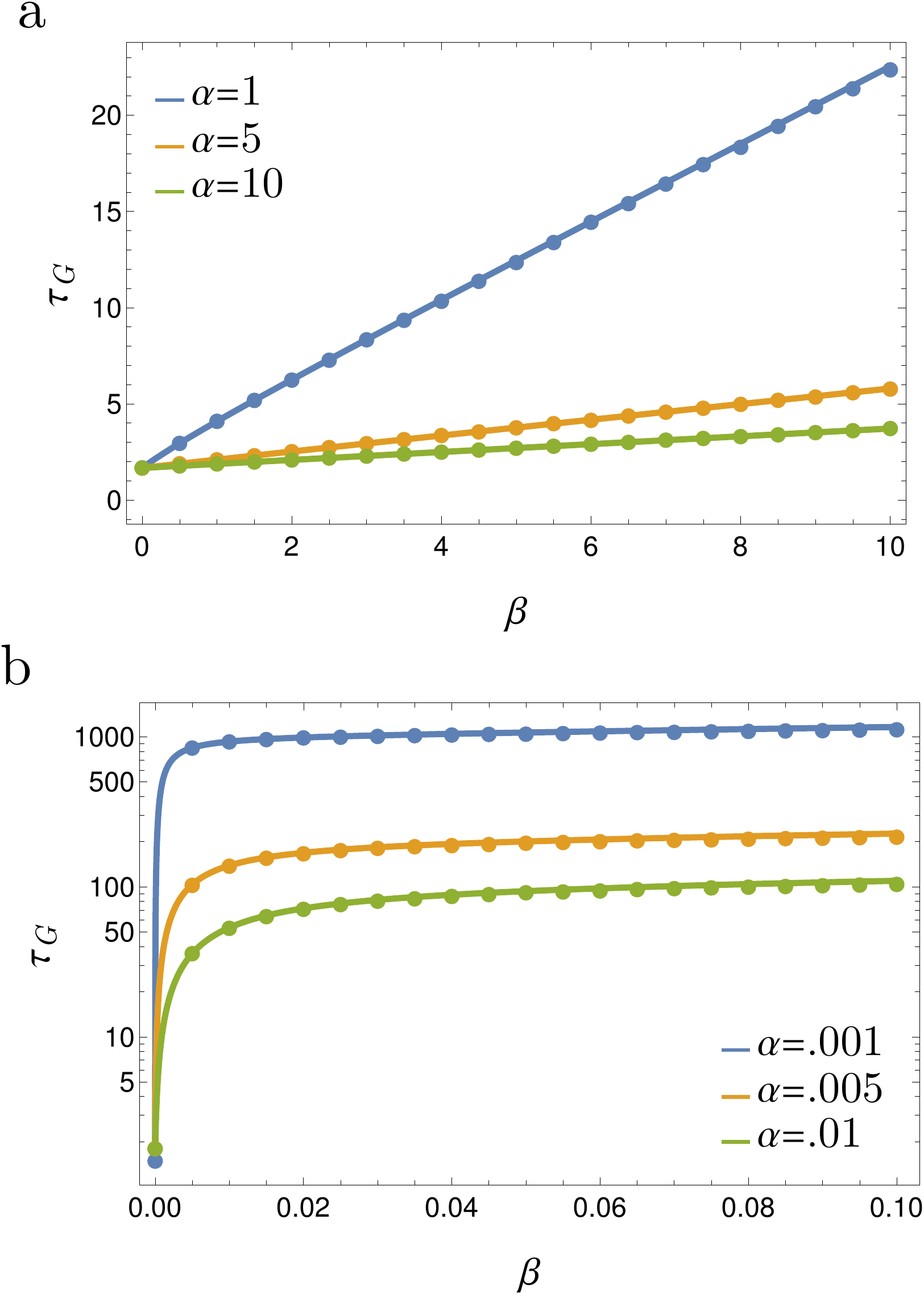}

			\caption{ Global mean first hitting time as a function of $\beta$ for $\ell=1$. a) $\alpha=\{1,5,10\}$ and b) $\alpha=\{.001,.005,.01\}$. Symbols represent the results from an average of $50,000$ Gillespie simulations.} 
\label{fig:tgforbeta}
\end{figure}

The global MFPT is split into two contributions: one part corresponding to the case of a perfectly absorbing target (setting $\beta=0$), and another which depends on both the parameters of the RTP (through $\ell=L\gamma/v$) and on the re-scaled target switching rates. In fact, one can see that $\tau_1(z)$ and $\tau_G$ are strictly increasing with $\beta$: the longer the target is found in the hidden state, the longer it takes for the RTP to be absorbed. Figures \ref{fig:tgforbeta}a-b depict the mean global time given by Eq. \eqref{GMeanTimeDimless} as a function of $\beta$ for several values of $\alpha$. It is interesting to notice that due to the hyperbolic function, the global MFPT abruptly increase at small $\beta$, as depicted in figure \ref{fig:tgforbeta}b. We successfully compare the analytic results with simulations that use the Gillespie algorithm \cite{GILLESPIE1976403}.

Defining 
\begin{equation}
    \epsilon=\frac{\alpha}{\alpha+\beta+\beta\sqrt{\frac{\alpha+\beta+2}{\alpha+\beta}}\coth{\ell\sqrt{\alpha+\beta}\sqrt{\alpha+\beta+2}}},\label{absorptionCoeff}
\end{equation}
equation \eqref{MeanTimeDimless} can be recast as
\begin{equation}
    \tau_G=\frac{\ell}{\epsilon}+\frac{2}{3}\ell^2.\label{GMeanTimeDimless2}
\end{equation}
With this notation, we notice that $\tau_G$ takes the same expression as for a RTP confined on one side by a partially reflecting boundary at the origin, with absorption coefficient $\epsilon$, and on the other side by a totally reflecting boundary placed at $\ell$\cite{angelani2015run}. In such model there is no target dynamics: at each passage at the origin, the RTP is absorbed with probability $\eta\equiv 2\epsilon/(1+\epsilon)$ and reflected with probability $1-\eta$\cite{angelani2015run,MasoliverPRE1993}.  In our problem, yet, due to the temporal variations of the target state, $\eta$ is not simply given by $\alpha/(\alpha+\beta)$, the probability that the target is found in the visible state, but depends in a more intricate way on the target rates and the effective turning rate $\ell$. The effective absorption probability in our problem is thus affected by the relative time scales associated with target switching (with respect to the turning time of the RTP) and by the relative time needed by a ballistic particle to cross the whole domain ($\ell$). 

In physical units, the global MFHT is given by 
\begin{equation}
t_G=\gamma^{-1}\tau_G,  
\end{equation}
and can be expressed from Eqs. \eqref{absorptionCoeff}-\eqref{GMeanTimeDimless2} using the dimensional parameters as
\begin{equation}
        t_G=\frac{L}{\kappa}+\frac{2L^2\gamma}{3v^2}.\label{GMeanTime}
 \end{equation}
 where $\kappa=v\epsilon$ is given by
\begin{equation}
    \kappa=\frac{va}{a+b+b\sqrt{\frac{a+b+2\gamma}{a+b}}\coth{\frac{L\sqrt{a+b}\sqrt{a+b+2\gamma}}{v}}}.\label{absorptionCoeffk}
\end{equation}

Several limiting cases ought to be mentioned:

{\it (i) Perfect absorption, $\beta=0$:} From Eq. \eqref{absorptionCoeff}, one obtains $\epsilon=1$, {\it i.e.}, the perfect absorption boundary condition. Hence, we recover the global mean first passage time
 \begin{equation}
       \tau_G(\beta=0)=\ell+\frac{2}{3}\ell^2,\label{tauGbeta0}
 \end{equation}
which was previously derived in \cite{angelani2014first}. 

{\it (ii) High transition rates $\alpha \gg 1$ and $\beta \gg 1$: }  In this scenario, the target transitions are so fast that the RTP only \lq\lq sees\rq\rq\  a partial absorbing boundary with an absorption coefficient $\epsilon\approx\frac{\alpha}{\alpha+2\beta}$. This leads to $\eta\approx\alpha/(\alpha+\beta)$, \textit{i.e.}, the probability that the particle is absorbed is equal to the probability that the target is found in the visible state. The global mean first passage time reads
 \begin{equation}
       \tau_G\approx\frac{\alpha+2\beta}{\alpha}\ell+\frac{2}{3}\ell^2.
 \end{equation}



{\it (iii) Low transition rates $\alpha \ll 1$ and $\beta \ll 1$: } In this case the absorption coefficient approaches
\begin{equation}
    \epsilon\approx\frac{\alpha}{\alpha+\beta+\beta\sqrt{\frac{2}{\alpha+\beta}}\coth{\ell\sqrt{2}\sqrt{\alpha+\beta}}}.
\end{equation}

{\it (iv) Ballistic  particle $\gamma=0$: }\label{deterministicsearcher} The other parameters being fixed, from Eq. \eqref{GMeanTime} one can notice that $t_G$ reaches a minimum at $\gamma=0$ and
 \begin{equation}
    t_G(\gamma=0)=\left(\frac{a+b}{a v}+\frac{b}{a v}\coth{\frac{L(a+b)}{v}}\right)L.\label{GMeanTimeBallistic}
\end{equation}
As in the stationary target case, the optimal strategy to react quickly with the dynamical target is ballistic motion. In this scenario, the particle performs straight line movements and only flips its direction when reflecting at the walls, crossing the origin periodically until it coincides with the target in the active state. On average, less persistent searchers waste time in fruitless excursions, not returning to the origin often enough to detect the target. In Appendix \ref{Ballistic} we deduce Eq. \eqref{GMeanTimeBallistic} by another method, from purely probabilistic arguments. 

{\it (v)  Brownian limit, $v\to\infty$, $\gamma\to\infty$ and $v^2/\gamma$ fixed:} Taking the limit of large $v$ and $\gamma$ with  $v^2/(2\gamma)\equiv D$, the stochastic noise $\Gamma(t)$ in Eq. \eqref{LangevinRT} becomes a white noise, leading to a Brownian motion with diffusion constant $D$\cite{Malakar_2018,singh2020run,Abhishek2019}.
 In this limit, Eq. (\ref{GMeanTime}) becomes\cite{szabo1980first}
\begin{equation}
   t_G=\frac{L}{\kappa}+\frac{L^2}{3D},\label{TGBrownian}
\end{equation}
with the effective reactivity coefficient given by
\begin{equation}
    \kappa=\frac{a}{b}\sqrt{D(a+b)}\tanh{L\sqrt{\frac{a+b}{D}}}.\label{absorptionCoeffk2}
\end{equation}
Eq. \eqref{TGBrownian} coincides with the expression for a bounded Brownian particle with a partially absorbing boundary on one side, obeying the Robin boundary condition:
\begin{equation}
    D \frac{d\rho}{dx}\Big|_{x=0}=\kappa \rho(x=0)
\end{equation}
where $\rho$ is the particle probability density.
 We remark that, when $L\longrightarrow\infty$, one recovers the reactivity coefficient $\kappa=\frac{a}{b}\sqrt{D(a+b)}$ that was first deduced in \cite{PRLFirstHittingTimes} for an unbounded Brownian particle. Therefore, Eq. \eqref{absorptionCoeffk2} generalizes the connection that exists between partially absorbing and intermittent boundaries for Brownian particles.



\section{Coefficient of variation}\label{SecRelVar}

The Laplace transform of the FHTD given by Eq. \eqref{meanP2} seem rather difficult to invert, however we can obtain from this expression the second moment of the distribution. Let us define the coefficient of variation of the first hitting time as
\begin{equation}
C_v=\frac{\langle\overline{[\tau'-\tau_G]^2}\rangle_z}{\tau_G^2},
\end{equation}
where the average $\langle\overline{\cdot}\rangle_z$ runs over both the realizations of the process and the starting position $z$. We use this quantity to assess the global fluctuations of $\tau'$ around its global mean, re-scaled by $\tau_G^2$. Therefore,
\begin{equation}
    C_v=\frac{\frac{1}{\ell}\int_0^{\ell} dz\int_0^{\infty} d\tau' P_{av}(z,\tau')(\tau'-\tau_G)^2}{\tau^2_G}
\end{equation}
 Integrating by parts and using the relation \eqref{fhtd}, the coefficient of variation can be written as
\begin{equation}
    C_v=-\frac{2}{\ell\tau_G^2 }\int_0^{\ell} dz \frac{\partial \widetilde{Q}_{av}(z,s)}{\partial s}\big|_{s=0}-1.
\end{equation}
Using Eq. \eqref{meanQ2}, one obtains 
\begin{align}
         C_v=& 1+\frac{2\ell^2}{\tau^2_G}\Bigg[ \frac{4\ell^2}{45}-\frac{1}{3}\nonumber\\
         &+ \frac{\beta(1+\alpha+\beta)}{\alpha (\alpha+\beta)}\left(
   \text{csch}^2X+\frac{\coth X}{(1+\alpha+\beta)X}\right)\Bigg],\label{relativeVarianceRT}
\end{align}
where $X$ is defined as
\begin{equation}
    X=\ell\sqrt{\alpha+\beta}\sqrt{\alpha+\beta+2}.\label{XforRTDimless}
\end{equation}

In the limit $\beta\to\infty$, $\tau_G$ diverges (the target is always invisible) and $C_v\to 1$, whereas for $\beta=0$ (the target is always visible) the coefficient of variation only depends on $\ell$. From Eq. \eqref{tauGbeta0} one gets
\begin{equation}
    C_v(\beta=0)=\frac{15+4\ell(15+7\ell)}{5(3+2\ell)^2}.\label{cvbeta0}
\end{equation}
To the best of our knowledge, this expression has not been derived in the literature on RTPs.

\begin{figure}[htp]
\centering
			\includegraphics[width=.48\textwidth]{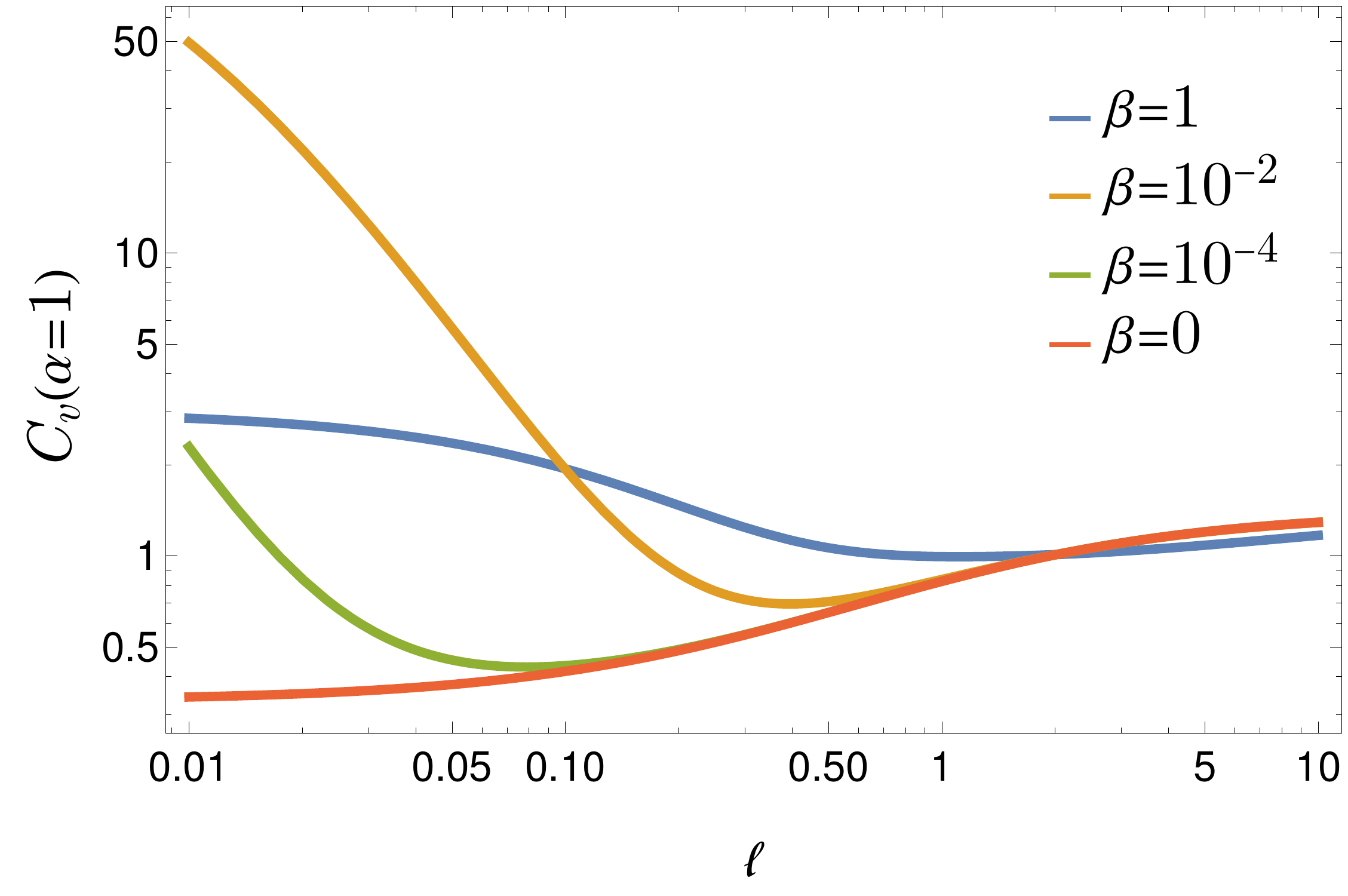}
            
			\caption{Relative variance as a function of $\ell$ for several values of $\beta$ at fixed $\alpha=1$. Symbols represent simulation results obtained with the Gillespie algorithm.}
			\label{fig:RelativeVarianceL}
\end{figure}

Eq. \eqref{cvbeta0} tells us that $C_v(\beta=0)$ increases monotonically with the dimensionless length $\ell$, which can be varied by moving the reflective walls further apart, or by changing the particle velocity $v$, or the turning rate $\gamma$. The coefficient of variation for a non-gated target is restricted to the interval $1/3\le C_v(\beta=0,\ell)<7/5$ (see Fig. \ref{fig:RelativeVarianceL}). However, for the more general case with $\beta>0$ and $\alpha>0$, the full expression of the coefficient of variation suggests a more intricate dependence with $\ell$. From numerical evaluations of Eq. \eqref{relativeVarianceRT} we observe that $C_v$ can take values $\gg 1$ and also reaches a minimum at a non-trivial length $\ell^*(\alpha,\beta)$, as depicted in Fig. \ref{fig:RelativeVarianceL}. Given $\alpha$ and $\beta$, the searcher can thus minimise the uncertainty on the first hitting time by adjusting its velocity to reach $\ell^*(\alpha,\beta)$.

\begin{figure}[hpt]
\centering
			\includegraphics[width=.48\textwidth]{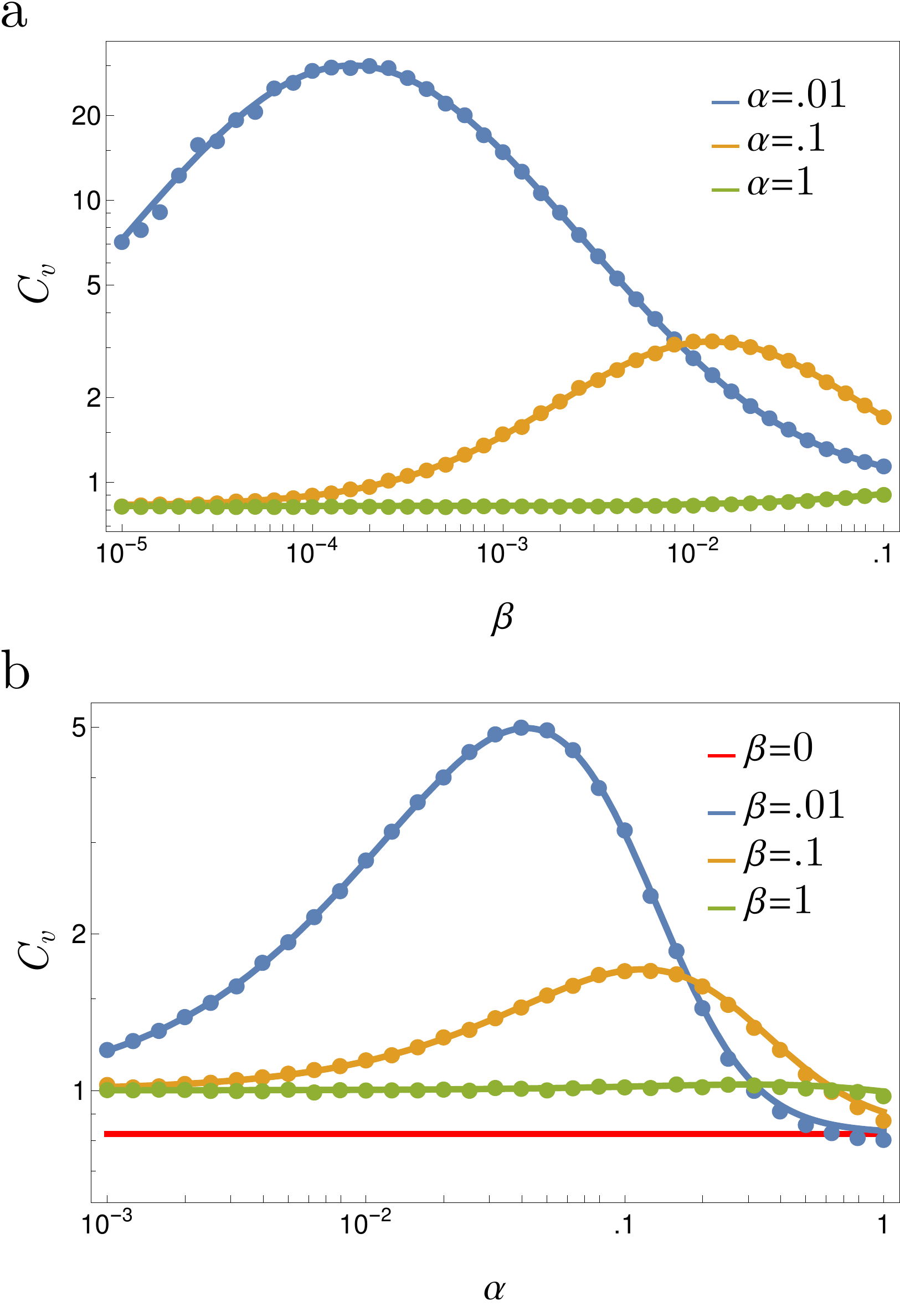}
			\caption{a) Coefficient of variation as a function of $\beta$ for several values of $\alpha$ at fixed $\ell=1$, b) Same quantity as a function of $\alpha$ for several values of $\beta$. In both cases, $\ell=1$. Symbols represent simulation results obtained with the Gillespie algorithm.}
			\label{fig:RelativeVariance}
\end{figure}

The curves of Fig. \ref{fig:RelativeVarianceL} also show a rather unexpected behaviour of the coefficient of variation: at fixed $\ell$ and $\alpha$, $C_v$ varies non-monotonically with $\beta$. As shown in Figures \ref{fig:RelativeVariance}a-b and numerical evaluations of Eq. \eqref{relativeVarianceRT}, $C_v$ reaches a maximum as the switching rates $\beta$ or $\alpha$ are varied at fixed $\ell$. This maximum indicates that the distribution suddenly widens around its mean for a particular value of the target rate. In those situations, the first hitting times become less predictable and the MFHT less meaningful. Notably, the coefficient of variation can reach values much larger than unity when the re-scaled rates become small  $(\alpha,\beta\ll 1)$. Fixing $\alpha\ll1$ and $\ell=1$, we observe in Fig. \ref{fig:RelativeVariance}a that $C_v$ peaks at a value $\beta^*$ which is much smaller than  $\alpha$, {\it i.e.}, when the inactive phases of the target are relatively brief compared to the active phases. This finding is paradoxical: the reaction time becomes widely unpredictable due to the target dynamics, but the target is most of the time reactive! On the other hand, when one fixes $\beta\ll 1$ and $\ell=1$, one finds that $C_v$ peaks at a value $\alpha^*$ which is of the same order as $\beta$ (see Fig.\ref{fig:RelativeVariance}b), that is, when the target spends on average the same time in the active phase as in the inactive phase.

\section{Infinite domain}\label{InfiniteDomain}

In this section, we study the RTP and the intermittent target in the limit of infinite domain size. In this case, the asymptotic forms of the survival probability can be obtained explicitly. For convenience, in the following analysis we will work with the dimensional variables $\{x,t,\gamma,a,b\}$ and use the Laplace transform $\widetilde{Q}(x,u)=\int^\infty_0e^{-ut}Q(x,t) dt$, which is equivalent to take $\widetilde{Q}(x,u)=\gamma^{-1}\widetilde{Q}(z,s)$ in Eq. \eqref{meanQ2}, with  $u=s\gamma$. By taking the limit $\ell\to \infty$ in Eq. (\ref{meanQ2}), we get
 \begin{equation}
     \widetilde{Q}_{av}(x,u)=\frac{1}{u}-\frac{e^{-\frac{x}{v}\sqrt{u(u+2\gamma)}}}{u+\frac{u}{a}\sqrt{\frac{u}{u+2\gamma}}\left(a+b+b\sqrt{\frac{u+a+b+2\gamma}{u+a+b}}\right)}.\label{Linf}
 \end{equation}

Despite the fact that this expression has not a simple form allowing an exact inversion from the Laplace domain, it is possible to analyze the long time behaviour of the survival probability and the associated first hitting time distribution. The long time regime can be extracted from the small $u$ expansion of the image function $\widetilde{Q}_{av}(x,u)$. Making the approximations $\sqrt{u(u+2\gamma)}\approx \sqrt{2\gamma u}$, $\sqrt{u/(u+2\gamma)}\approx\sqrt{u/(2\gamma)}$ and   $\sqrt{a+b+u}\approx \sqrt{a+b}$, Eq. (\ref{Linf}) becomes
 \begin{equation}
     \widetilde{Q}_{av}(x,u)\simeq\frac{1}{u}\left(1-\frac{e^{-\frac{x}{v}\sqrt{2\gamma u}}}{1+R\sqrt{u}}\right)\label{approxQavRT}
 \end{equation}
 where we have defined $R=\frac{1}{\sqrt{2\gamma}a}\left(a+b+b\sqrt{\frac{a+b+2\gamma}{a+b}}\right)$.
     The above expression can be exactly inverted \cite{abramowitz1965handbook} to yield
 \begin{align}
    Q_{av}(x,t)\simeq&\operatorname{erfc}\left(\frac{\sqrt{t}}{R}+\frac{x}{v}\sqrt{\frac{\gamma}{2t}}\right)\exp\left(\frac{x\sqrt{2\gamma}}{vR}+\frac{t}{R^2}\right)\nonumber\\
    &+\operatorname{erf}\left(\frac{x}{v}\sqrt{\frac{\gamma}{2t}}\right),\label{QavLargeTRT}
\end{align}
and, from Eq. \eqref{fhtd},
 \begin{align}
    P_{av}(x,t)\simeq&\frac{1}{R\sqrt{\pi t}}\exp\left(-\frac{\gamma x^2}{2tv^2}\right)\nonumber\\
    &-\frac{1}{R^2}\operatorname{erfc}\left(\frac{\sqrt{t}}{R}+\frac{x}{v}\sqrt{\frac{\gamma}{2t}}\right)\exp\left(\frac{x\sqrt{2\gamma}}{vR}+\frac{t}{R^2}\right),
    \label{PavLargeTRT}
\end{align}
where  $\operatorname{erf}(z)=\frac{2}{\sqrt{\pi}}\int_0^z e^{-\xi^2}d\xi$ is the error function and $\operatorname{erfc}(z)=1-\operatorname{erf}(z)$ the complementary error function. Due to the approximations made, Eqs. \eqref{QavLargeTRT}-\eqref{PavLargeTRT} hold for $t$ larger than both the target relaxation time $t_{ta}\equiv (a+b)^{-1}$ and the tumble time $t_{tb}\equiv (2\gamma)^{-1}$. 

From the above equations one can see that the first hitting time distribution (and also the survival probability) is determined by two characteristic timescales: a diffusive time $t_{D}=x^2\gamma/(2v^2)$, which is the typical time needed for the particle to reach the origin, and the time 
\begin{equation}
t_c=R^2= \frac{1}{2\gamma a^2}\left(a+b+b\sqrt{\frac{a+b+2\gamma}{a+b}}\right)^2\label{crossovertc},
\end{equation}
that sets a crossover time that separates two different scaling regimes in the asymptotic behaviour of the FHTD. These regimes are deduced from Eq. \eqref{PavLargeTRT} as follows (the analysis can also be done directly from the image function $\widetilde{Q}_{av}(x,u)$ in Eq. \eqref{approxQavRT}, see \cite{PRLFirstHittingTimes}):

\begin{itemize}
    
\item {\it The true asymptotic limit $t\gg t_{c}$:} in this limit we can use the approximation $\operatorname{erfc}(x)\approx \frac{e^{-x^2}}{\sqrt{\pi}x}\left(1-\frac{1}{2x^2}+\dots\right)$ to get 
    \begin{equation}
P_{av}(x,t)\simeq \frac{x\sqrt{2\gamma}/v+R}{2\sqrt{\pi t^3}},\  \ t\gg t_c,
    \end{equation}
    which is the tail of the L\'evy-Smirnov distribution typical of Brownian motion and random walks \cite{feller2008introduction}, but with a different prefactor. The modification of this prefactor is a phenomenon that has also been observed in random search problems with fluctuating targets on networks, including the case of non-Markovian switching dynamics \cite{caceres1995theory,Budde_1995}.

\item{\it The intermediate regime:}
if $t_{c}$ is much larger than all the other characteristic times, or $t_{c}\gg {\rm max}(t_{D},t_{ta},t_{tb})$, the arguments in the exponential and the complementary error functions are small and we notice that, before the true asymptotic regime, an intermediate time regime appears:
\begin{equation}\label{interreg}
    P_{av}(x,t)\simeq
\frac{1}{R\sqrt{\pi t}},\  \ \ {\rm max}(t_{D},t_{ta},t_{tb}) \ll t\ll t_{c}.
\end{equation}
\end{itemize}
Eq.(\ref{interreg}) represents a much slower decay than the standard $t^{-3/2}$ scaling. For simplicity,
in the following we will assume $t_{D}\ll t_{ta}, t_{tb}$, a condition which is easily enforced by choosing $x=0$, {\it i.e.}, by initially placing the particle right on the target. As we have done in Section \ref{MFHT}, we analyze below different limiting cases.

{\it (i) Perfect absorption, $b=0$:} From Eq. \eqref{crossovertc} one obtains $t_{c}=t_{tb}$, therefore the conditions for the existence of the intermediate regime are not fulfilled and we find the asymptotic decay\cite{Malakar_2018}
\begin{equation}
P_{av}(x,t)\simeq \left(x\sqrt{2\gamma}/v+1/\sqrt{2\gamma} \right)/\sqrt{4\pi t^3}.
\end{equation}

{\it (ii) High transition rates $a,b\gg \gamma$: } In this case ${\rm max}(t_{ta},t_{tb})=t_{tb}$ and $t_{c}\simeq\left(\frac{a+2b}{\sqrt{2\gamma}a}\right)^2$, therefore 
\begin{equation}
    \frac{t_{c}}{t_{tb}}\simeq\left(\frac{a+2b}{a}\right)^2.\label{crossoverHighrates}
\end{equation}

{\it (iii) Low transition rates $a,b \ll \gamma$: } Here, ${\rm max}(t_{ta},t_{tb})=t_{ta}$ and $t_{c}\simeq\frac{b^2}{a^2(a+b)}$, then
\begin{equation}
    \frac{t_{c}}{t_{ta}}\simeq \frac{b^2}{a^2}.\label{crossoverLowrates}
\end{equation}

{\it (iv) Ballistic  particle $\gamma=0$: } From the inversion of Eq. \eqref{Linf}, one obtains that, for  the ballistic searcher, $P_{av}(x,t)=\frac{a\delta(t-x/v)}{2(a+b)}$ for $t>0$, which is the condition for the particle to start at $x>0$ and move, with probability $1/2$, in a straight line towards the origin and then, with probability $a/(a+b)$, hit the target.

{\it (v)  Brownian limit, $v\to\infty$, $\gamma\to\infty$ and $v^2/\gamma$ fixed:} In this case we recover the result
\begin{equation}
\frac{t_{c}}{t_{ta}}=\frac{b^2}{a^2}\label{crossoverBrownian}
\end{equation}
deduced in \cite{PRLFirstHittingTimes} for a Brownian particle, and which actually coincides with Eq. \eqref{crossoverLowrates}.

\begin{figure}[htbp]
\centering
      \includegraphics[width=.48\textwidth]{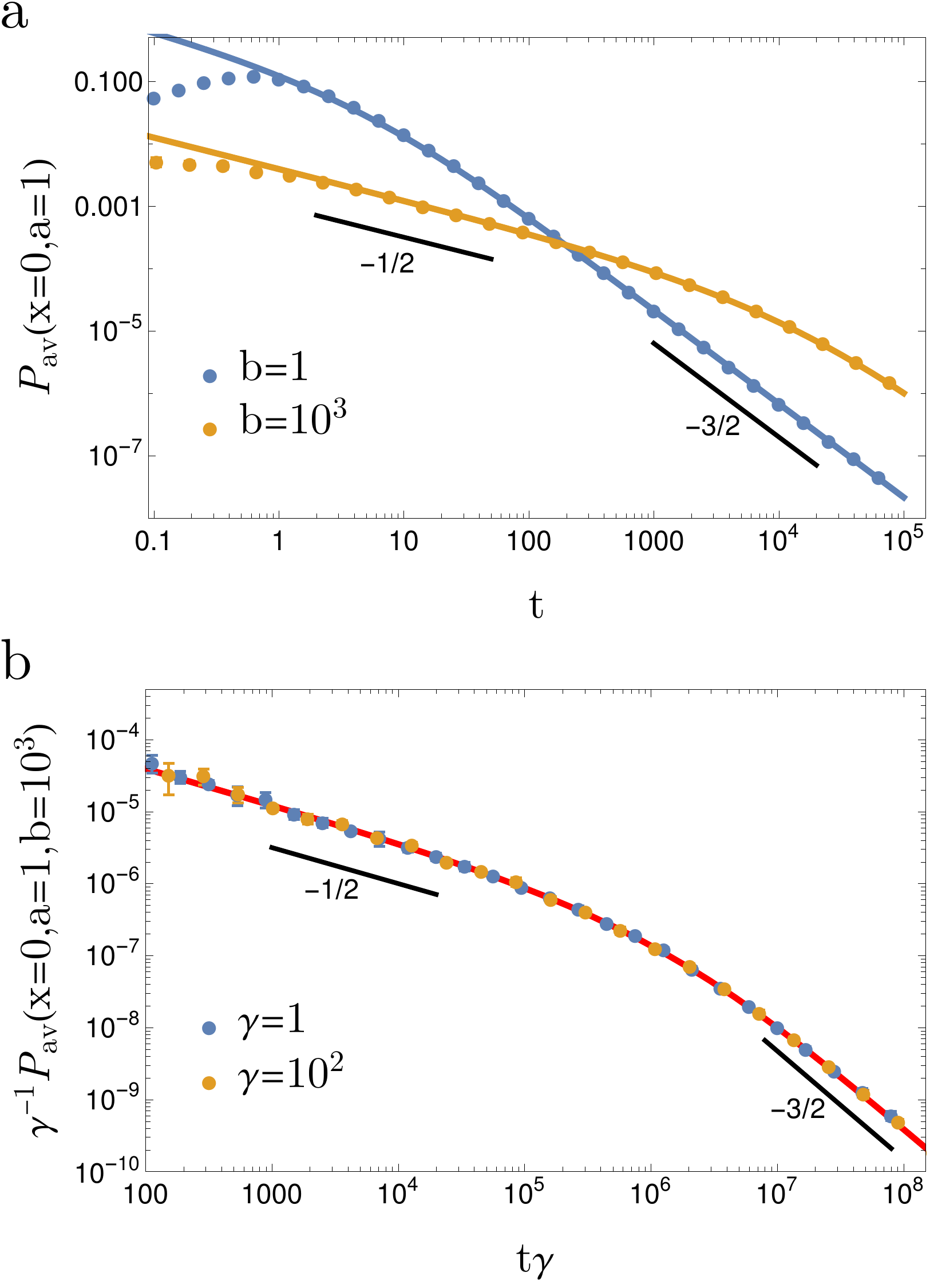}

			\caption{a) FHTD for a run-and-tumble particle that start at $x=0$, with $a=1$, $\gamma=1$ and 
			$b=\{1,10^3\}$. b) Same quantity for $b=10^3$ and $\gamma=\{1,10^2\}$. Lines are the analytical solution given by Eq. \eqref{PavLargeTRT} and symbols represents simulation results.}
\label{fig:meanfpt}
\end{figure}

From Eqs. \eqref{crossoverHighrates}-\eqref{crossoverBrownian}, we conclude that the intermediate scaling regime $\sim t^{-1/2}$ of the FHTD  emerges when the target is cryptic, \textit{i.e.}, when $b\gg a$ (see Fig. \ref{fig:meanfpt}a). We also see that the crossover time becomes longer as $\beta$ increases. 
In addition to this, the result in Eq. \eqref{crossoverBrownian} is interesting since it implies that the Brownian motion becomes the best strategy that the searcher can opt in order to reduce (as much as possible) the slow decay range ($\sim t^{-1/2}$) in the FHTD. This is owing to the recurrence property of the Brownian motion, that ensures that the particle recrosses the origin many times within a short period of time. 

\section{Discussion}\label{Discussion}
We have studied the first hitting time statistics between a run-and-tumble particle and a target that randomly switches between two states, visible and invisible. Target detection is allowed only in the visible state. When the system is confined between two reflective walls, we have focused our attention on the mean first hitting time, which can be calculated exactly. For any values of the rates $a$ and $b$ of the intermittent dynamics, the particle should opt for a ballistic motion ($\gamma=0$) in order to minimize the mean search time. Less persistent motion not only increases, on average, the time of first target encounter, but also makes this time less predictable, \textit{i.e.}, with larger relative fluctuations around the mean. According to the results of section \ref{SecRelVar}, the relative variance of the first hitting time exhibits a non-monotonic behaviour with respect to the intermittent parameters and the turning rate. When the transition rates become slow compared to the tumbling rate $(a,b\ll\gamma)$, the coefficient of variation takes larger values compared with the steady target case.

When the particle motion is unbounded, our findings extend the results of \cite{PRLFirstHittingTimes} on Brownian motion and an intermittent target on the infinite line. We have found that the target dynamics drastically affect the scaling of the first hitting times distribution, whose most unusual feature is an extended intermediate regime in $t^{-1/2}$, previous to the standard $t^{-3/2}$ asymptotic decay. The crossover time $t_c$ that separates the two regimes depends on the rates $\gamma$, $a$ and $b$, and is given by Eq. \eqref{crossovertc}. On time scales smaller than $t_c$, the search process is slow in the sense that the survival probability decays much more slowly than in the standard case of a perfectly reactive target. 


The expression of the global MFPT in bounded domains in our problem takes the same form as for the RT with partial absorption, although the two problems are clearly different. 
Particularly interesting is to notice that the crossover time $t_c$ can be recast under a simple form as $t_c=2\gamma\epsilon^{-2}$, where $\epsilon$ is the absorption coefficient given by Eq. \eqref{absorptionCoeff} it the limit $\ell=\infty$. Therefore, the crossover time can be redefined in terms of the tumble rate and the effective absorption coefficient. As soon as $t$ is $\gg$ than $t_{ta}$ and $t_{tb}$ (the target relaxation and particle turning times, respectively), the scaling behaviour of the first hitting time distribution and all the subsequent conclusions remain valid for the problem of partial absorption. A similar correspondence has been already pointed out for Brownian diffusion in recent studies\cite{grebenkov2019imperfect,PRLFirstHittingTimes}. Our study brings further understanding on non-equilibrium particles diffusing in fluctuating environments, where target encounters depend on both internal noises and fluctuations external to the particle dynamics.


{\bf Acknowledgements:} 
GMV thanks CONACYT for a scholarship support.

\appendix
\section{Backward Fokker-Planck equations}\label{App_BFP}

To derive the set of equations \eqref{SPsystem1}, let us first suppose that at time $t=0$ the particle starts with velocity $+v$ and the target is in the invisible state or $\sigma(t=0)=0$. During the small time interval $[0,\Delta t]$ there is a probability $a\Delta t$ for the target to change to the state $\sigma=1$ and a probability $1-a\Delta t$ to remain in the state $\sigma=0$. On the other hand, with a probability $\gamma\Delta t$ the particle will change its velocity to $-v$, or will remain with $+v$ with probability $1-\gamma\Delta t$. If we sum these contributions, the survival probability can be written as 
\begin{equation}
\begin{aligned}
 	Q^+_0(x,t+\Delta t)=&(1-a\Delta t)(1-\gamma\Delta t)Q^+_0(x+v\Delta t,t)\\
 	&+a\Delta t(1-\gamma\Delta t) Q^+_1(x+v\Delta t,t)\\
 	&+\gamma \Delta t(1-a\Delta t) Q^-_0(x+v\Delta t,t)\\
 	&+a\gamma\Delta t^2Q^-_1(x+v\Delta t,t).\label{Q0plusChapman}
 \end{aligned}
 \end{equation}
Expanding the r.h.s. of \eqref{Q0plusChapman} in Taylor series and retaining only the terms of order $\Delta t$, we obtain
\begin{equation}
    \frac{\partial Q^+_0}{\partial t}=v\frac{\partial Q^+_0}{\partial x}-a(Q_0^+-Q_1^+)-\gamma(Q_0^+-Q_0^-),
\end{equation}
which is the first equation in \eqref{SPsystem1}. The other three equations for $Q_0^-$, $Q_1^+$ and $Q_1^-$ can be deduced in a similar way.

\section{MFHT for a ballistic particle}\label{Ballistic}

 In the ballistic limit $\gamma=0$, the motion of the RTP becomes deterministic; once the particle starts moving, it will cross the origin periodically, with period $T=2L/v$, until it coincides with the target in the visible state. Then, in order to compute the survival probability it is enough to find the probability that the target is hidden at a succession of times periodically spaced. We also set $x=0$, as we can see from Eq. \eqref{MeanTimeDimless} for a searcher with $\gamma>0$ that $t_G=t(x=0)$. For a ballistic particle starting at $x>0$, the first crossing of the origin occurs at time $x/v$ for the initial velocity $-v$ and $(2L-x)/v$ for the initial velocity $+v$, which gives the average time of $2L/v$, independent of $x$. We therefore choose $x=0$. With this initial condition the contribution to the survival probability for the initial velocity $-v$ is zero when the target initial state is $\sigma_0=1$, \textit{i.e.}, $Q^-_1=0$. Also by symmetry one has that $Q^+_0=Q^-_0$. In the following we focus on the calculation of $Q^+_0$ and $Q^+_1$. 

Let us consider a Poisson point process $\Pi_{a+b}$ with rate $a+b$ in dimensional units. The probability that an $a$-event is followed by a $b$-event is $p=a/(a+b)$, whereas $q=b/(a+b)$ is the probability that the $b$-event is followed by an $a$-event. We can take each case as a Bernoulli realization with probability of success $p$ and failure $q$, and the number of realizations will be distributed in time as $\Pi_{a+b}$. The probability that at time $t$ the target is invisible given the initial target state $\sigma(t=0)=\sigma_0$ is
\begin{equation}
    P\left[\sigma(t)=0|\sigma_0\right]=\frac{b}{a+b}\left(1+C_{\sigma_0} e^{-(a+b)t}\right),\label{ProbSigma0}
\end{equation}
where $C_0=a/b$ and $C_1=-1$. For $\sigma_0=0$, Eq. \eqref{ProbSigma0} is the probability that the last event occurred is not $a$, whereas for $\sigma_0=1$ it is the probability that the last event occurred is $b$.

Similarly, the probability that the target is visible at time $t$ given the initial state $\sigma_0$ is
\begin{equation}
    P\left[\sigma(t)=1|\sigma_0\right]=\frac{a}{a+b}\left(1+B_{\sigma_0} e^{-(a+b)t}\right)\label{ProbSigma1}
\end{equation}
where $B_0=-1$ and $B_1=a/b$.

As mentioned, the searcher will cross the origin at times $t_n=nT$ for $n\in\{1,2,\dots\}$ and from Eq. \eqref{ProbSigma0} the survival probability at time $t_n$ given the initial velocity $+v$ and the initial target state $\sigma_0=0$ is $P\left[\sigma(t)=0|\sigma_0=0\right]^n$ or  
\begin{equation}
Q_0^{+}=\left(\frac{b}{a+b}+\frac{a e^{-(a+b)T}}{a+b}\right)^{n}
\end{equation}
For the initial condition $\sigma_0=1$ one has
\begin{equation}
Q_1^{+}=\frac{b}{a+b}\left(1- e^{-(a+b)T}\right)\left(\frac{b}{a+b}+\frac{a e^{-(a+b)T}}{a+b}\right)^{n-1}
\end{equation}
which is the probability that the target has switched to invisible at the first passage of the particle and is invisible at the following $n-1$ passages.

Averaging over the initial velocities and target initial conditions, with $Q^-_1=0$ and $Q^+_0=Q^-_0$, the average survival probability up to time $t_n=nT$ will be
\begin{equation}
\begin{aligned}
    Q_{av}(n)&=\frac{ab}{(a+b)^2}\left[\frac{b}{a}+e^{-(a+b)T/2}\cosh{\frac{(a+b)T}{2}}\right]\\
    &\times\left(\frac{b}{a+b}+\frac{a e^{-(a+b)T}}{a+b}\right)^{n-1}\label{QavProb}
\end{aligned}
\end{equation}
where $n\in\{1,2,\dots\}$. 

As expected, for $n=0$ the survival probability is $Q_{av}(n=0)=\frac{b}{a+b}+\frac{a}{2(a+b)}$, since there is a probability $b/(a+b)$ that at time $t=0$ the target is invisible, and a probability $a/(2(a+b))$ that at time $t=0$ the target is visible but the searcher starts moving with velocity $+v$. 

From Eq. \eqref{QavProb}, the global MFHT will be
\begin{equation}
\begin{aligned}
    t_G&=\sum_{n=0}^\infty t_nP_{av}(t_n)=T\sum_{n=0}^\infty nP_{av}(n)=T\sum_{n=0}^\infty Q_{av}(n)\\
    &=\left(a+b+b\coth{\frac{L(a+b)}{v}}\right)\frac{L}{a v},
\end{aligned}
\end{equation}
and Eq. \eqref{GMeanTimeBallistic} is recovered.

\bibliographystyle{apsrev4-1}
\bibliography{Biblio}

\end{document}